# Popularity Without Legitimacy? Comparing Trust in Television Meteorologists and YouTube Weatherfluencers

**Julie A. Vera**
Human Centered Design & Engineering
University of Washington, Seattle
jvera@uw.edu

**Mark Zachry**
Human Centered Design & Engineering
University of Washington, Seattle
zachry@uw.edu

**David W. McDonald**
Human Centered Design & Engineering
University of Washington, Seattle
dwmc@uw.edu

### ABSTRACT

During severe weather events, people must interpret rapidly evolving information to make time-sensitive safety decisions. Broadcast meteorologists have traditionally served as credentialed intermediaries within established media organizations, while independent "weatherfluencers" on YouTube have emerged as prominent real-time interpreters for large and growing audiences. This mixed-methods study (survey n=58; interviews n=8) provides one of the first empirical comparisons of how viewers evaluate broadcast meteorologists against YouTube weatherfluencers across credibility, legitimacy, objectivity, and practical utility. Broadcast meteorologists were consistently rated higher on credibility, legitimacy, and safety utility, while weatherfluencers achieved parity on objectivity. Yet weatherfluencer audiences continue to grow, revealing a critical decoupling between audience attention and official or professional authorization that existing crisis communication models do not fully account for. Qualitative findings illuminate the mechanisms underlying these judgments and their implications for emergency communication in hybrid information ecosystems.

#### Keywords

Digital Media, Livestreaming, Severe Weather, Weatherfluencers.

### INTRODUCTION

During severe weather events, individuals must interpret uncertain and evolving information to make time-sensitive safety decisions (Morss et al., 2017). Tornadoes, flash floods, hurricanes, and other extreme hazards demand communication that is not only accurate but also trusted, intelligible, and actionable under pressure.

Historically, severe weather communication in the United States has been organized around a formal infrastructure that includes official warning agencies such as the National Weather Service and broadcast meteorologists embedded in news organizations. This arrangement reflects a common premise in warning-response research: that established source credibility supports protective action, translating professional expertise and organizational accountability into attention, understanding, and compliance with warnings and guidance (Mileti & Sorensen, 1990; Lindell & Perry, 2012).

Independent digital creators have emerged as prominent interpreters of severe weather information, challenging the assumed alignment between professional authority and public attention. We refer to these creators as *weatherfluencers*, based on Hurcombe's characterization of the "newsfluencer" (Hurcombe, 2025). Weatherfluencers are livestreamers who perform real-time radar analysis, warning interpretation, and storm updates to live audiences on platforms such as YouTube, operating outside broadcast media and emergency management organizations and without the professional credentialing, editorial oversight, or formal accountability that characterize weather communication. In practice, weatherfluencer livestreams function as participatory sensemaking environments: creators narrate unfolding conditions continuously while simultaneously monitoring live chat through which viewers contribute local ground observations, ask interpretive questions, and seek guidance about specific locations. This participatory structure positions the weatherfluencer not simply as a broadcaster but as a real-time collaborative interpreter. Creators aggregate distributed local observations alongside radar data, producing a form of interpretation that differs structurally from the one-to-many communication model of traditional broadcast meteorology. Some weatherfluencers are trained meteorologists; others are not. Yet regardless of credential status, weatherfluencers are attracting and growing substantial audiences. The two creators examined in this study had approximately 2 million and 175,000 subscribers respectively at the time of data collection in April 2024; by early 2026, those figures had grown to approximately 3.1 million and 1.7 million. These are not marginal actors in the severe weather information environment. They are persistent and expanding presences that coexist with official and professional information sources, assembling audiences through platform recommendation systems, subscriber networks, and cross-platform communities rather than through the professional credentialing structures that have historically organized weather communication authority.

This growth raises a question that existing crisis communication frameworks are not well equipped to answer. Warning-response models often presume that professional actors' credibility supports message receipt and protective action (Mileti & Sorensen, 1990; Lindell & Perry, 2012). Crisis informatics research has examined how credibility judgments emerge through collective verification and sensemaking during crisis events (Mendoza et al., 2010; Starbird et al., 2014), but has placed less emphasis on the standing, authority, or legitimacy of the communicators interpreting risk for the public. Neither tradition fully addresses what happens when credentialed and platform-native interpreters coexist during the same hazard event, performing overlapping interpretive functions for audiences who must evaluate them simultaneously. If weatherfluencers are evaluated as less legitimate and less appropriate for safety-critical guidance, why do they continue to attract and grow audiences? And if their audiences are expanding, what does that imply for how we understand the relationship between historical legitimacy and audience attention in crisis communication?

To address these questions, we examine how viewers evaluate severe weather information delivered through two coexisting communication infrastructures: traditional broadcast television meteorology and YouTube-based weather livestreaming. We focus on how audiences differentiate evaluations of the information from evaluations of the communicator, and how these judgments connect to perceived usefulness for safety-relevant decision-making as conditions evolve. Accordingly, we ask:

RQ1: How do viewers compare broadcast meteorologists and YouTube weather livestreamers when evaluating trustworthiness, credibility, and legitimacy, and how do these judgments differ for the information versus the communicator?

RQ2: How do viewers make sense of severe weather information across broadcast and livestream formats as conditions evolve, particularly in terms of temporal updating, place-based interpretation, and perceived usefulness for safety decision-making?

By examining how broadcast meteorologists and weatherfluencers are evaluated relative to one another during unfolding severe weather events, we make visible a critical decoupling between the capacity to attract and sustain audience attention and the professional standing that has historically authorized safety-critical guidance. This decoupling has implications for crisis communication research, emergency management practice, and the growing population of platform-native communicators who interpret hazard information for large public audiences.

## LITERATURE REVIEW

Social media platforms have reshaped how crisis information is produced, circulated, and evaluated during disasters, supporting rapid dissemination, peer-to-peer communication, and emergent forms of situational awareness among affected publics and response organizations, while also introducing challenges such as uneven credibility, rumor propagation, and large-scale information flows that participants must interpret in real-time (Mendoza et al., 2010; Starbird et al., 2014). These literatures provide complementary but incomplete accounts of evaluation when multiple interpreters operate simultaneously across professional and platform-based environments.

### Severe Weather Risk Communication and Trust

Research on severe weather risk communication has long emphasized trust as a prerequisite for effective public response during hazards. In the United States, warning and response infrastructures have historically been organized around official and professional sources, including the National Weather Service (NWS), local emergency management agencies, and broadcast meteorologists embedded in legacy media organizations. Classic warning response models conceptualize trust and source credibility as relationships between publics and official warning systems that support attention, comprehension, and compliance with protective guidance (Mileti & Sorensen, 1990; Lindell & Perry, 2012). Complementing these models, Wachinger et al. (2013) demonstrate that trust in risk-related institutions is socially produced and historically grounded, shaped by prior experience with institutions, perceptions of legitimacy, and the broader credibility of warning systems over time.

Within this tradition, professional affiliation, professional credentialing, and organizational accountability function as durable signals that bundle perceived expertise with appropriate authority to issue safety-critical information. Weather-specific research further shows that trust in official and professional weather sources shapes how people interpret warnings, assess risk, and decide whether to take protective action (Sherman-Morris, 2005; Ripberger et al., 2015). Broadcast meteorologists have historically served as interpretive intermediaries, translating technical meteorological products into locally meaningful, place-based guidance through professional conventions, sustained community presence, and routine forecasting practices (Henson, 2013).

### Credibility Assessment in Crisis Informatics

Crisis informatics research approaches source evaluation from a different analytic starting point. Where warning-response research treats official and professional authority as a stable precondition for credibility, crisis informatics has examined how credibility judgments emerge dynamically through processes of community verification, correction, and collective sensemaking as information diffuses at scale (Palen et al., 2009; Mendoza et al., 2010; Starbird et al., 2014). This tradition has also documented ad hoc information hubs named after crisis events, such as social media pages created around the 2014 Carlton Complex Wildfire, that accumulate visibility despite uncertain operators and unverified credibility (Chauhan & Hughes, 2017, 2018).

These traditions emphasize different aspects of evaluation rather than entirely distinct constructs. Severe weather risk communication research has typically examined trust as a relationship between the public and warning systems that enables compliance, whereas crisis informatics has more often operationalized credibility as judgments about the quality or plausibility of specific pieces of information circulating across platforms. By centering content-level assessment and diffusion dynamics, crisis informatics has produced vital insights into rumor propagation and uncertainty management, but it has placed less emphasis on the standing, authority, or legitimacy of the communicators interpreting risk for the public.

### Platform-Based Communication and Crisis Interpretation

As severe weather communication increasingly unfolds across both formal and platform-mediated environments, publics may encounter multiple interpreters performing overlapping explanatory work. Crisis informatics research has documented emergent, event-specific information resources that accumulate visibility outside formal institutions. Chauhan and Hughes describe Crisis Named Resources (CNRs) as social media accounts and pages named after a crisis event that often become easy-to-find venues for information exchange, despite uncertainty about who operates them or how credible they are (Chauhan & Hughes, 2018). Subsequent work extended this finding to other events and platforms, including multi-platform COVID-19 CNRs spanning Facebook, Twitter, and Reddit, documenting the persistence of this pattern across hazard types and media environments (Chauhan & Hughes, 2018, 2020, 2021). A smaller but influential line of research in emergency management has begun to examine how organizations such as emergency management agencies and public information officers adapt to social media, navigate credibility challenges, and manage authority in public-facing online environments (Hughes & Palen, 2012; Hughes & Chauhan, 2015). This work foregrounds organizational practices and constraints, but offers limited insight into how audiences evaluate individual communicators during unfolding events.

Complementary research on online credibility from HCI and communication studies shows that audiences rely on visible cues and heuristics, such as presentation style, consistency, interactivity, and perceived expertise, when evaluating information and its source (Fogg, 2003; Metzger, 2007; Sundar, 2008). Platform-native news content creation introduces additional structural dynamics that existing credibility frameworks do not fully address. Hurcombe identifies *platformatization* as a defining feature of newsfluencers, noting that platform revenue streams are not incidental to their content production, but structurally embedded within it. Recommendation algorithms create pressure toward attention-maximizing presentation, while parasocial intimacy and relational labor characterize the audience relationships these creators cultivate, in place of the community accountability that grounds broadcast authority. These structural differences become consequential when formal weather communication infrastructure and platform-based weather communicators coexist during the same hazard event.

Creators without formal affiliation perform interpretive work similar to that of broadcast meteorologists, requiring audiences to navigate tensions among legitimacy, credibility, and trustworthiness. This study addresses that gap by examining how these evaluative judgments diverge and interact when viewers directly compare broadcast meteorologists and YouTube weatherfluencers during unfolding severe weather events.

## METHODS

We conducted a convergent, parallel mixed-methods design, combining quantitative survey assessments with qualitative semi-structured interviews to examine how viewers evaluated traditional television meteorologists against YouTube weatherfluencers.

### Study Design

The quantitative component used a within-subject design in which all participants evaluated all four weather sources, holding the meteorological event constant across clips to isolate platform and presenter effects. Data were collected through two parallel recruitment streams. First, we deployed a quantitative survey through Amazon Mechanical Turk (MTurk) to achieve breadth and statistical power. MTurk workers first completed a brief eligibility screener. Eligibility required participants to be 18 years or older, have prior experience with severe weather events, and report regular weather information consumption. Recruitment continued until 50 eligible participants completed the survey.

Second, we conducted in-depth interviews with eight additional participants recruited separately through Twitter/X, Bluesky, and Reddit. Interview participants first completed the same screener and survey instrument prior to participating in a one-hour think-aloud interview discussing their responses. These interview participants did not overlap with the MTurk sample. The final analytic survey sample included 58 responses (50 MTurk participants and 8 interview participants). Interview participants received $15 compensation for completing both the survey and the interview session. All study procedures were reviewed and approved by the authors' institutional review board.

### Participants

The final analytic survey sample (N = 58) consisted of adults aged 18 and older representing multiple age groups, with the largest share between 25–44 years old, followed by 45–54 years. Participants were distributed across U.S. regions, with strongest representation from the South and Midwest, followed by the West and Northeast. Respondents reported prior experience with severe weather including tornadoes, hurricanes, floods, and winter storms, and most indicated experiencing severe weather seasonally or monthly. Participants also reported consuming weather information through both local television broadcasts and YouTube-based weather content.

The interview sample included five women and three men. Based on self-reported behaviors, four participants were categorized as weather-knowledgeable (WK) and four as average weather-checkers (AWC).

**Table 1. Interview Participant Characteristics**

| Participant | Category | Gender | Residential Setting | Highest Education |
|---|---|---|---|---|
| P1 | AWC | F | Suburban | Master's |
| P2 | WK | M | Suburban | Master's |
| P3 | WK | F | Suburban | Bachelor's |
| P4 | AWC | F | Urban | Doctoral |
| P5 | AWC | M | Urban | Master's |
| P6 | WK | M | Suburban | Doctoral |
| P7 | AWC | NB | Rural | Master's |
| P8 | WK | F | Urban | Doctoral |

### Survey Instrument

Participants assessed all four weather sources using six dimensions adapted from Meyer's Credibility Index (Meyer, 1988), originally developed to operationalize newspaper credibility for readership research and subsequently extended across broadcast television, online news, and social media contexts. We retained Meyer's core dimensions while extending the instrument in one theoretically motivated direction by separating legitimacy into two analytically distinct items. Prior applications of Meyer-derived instruments typically treat legitimacy as a unitary construct (Gaziano & McGrath, 1986; Meyer, 1988).

The six dimensions were objectivity, accuracy, trustworthiness, credibility, coverage legitimacy, and presenter legitimacy. Each dimension included a brief embedded operational definition to support consistent interpretation: objectivity (reliance on scientific meteorological data rather than opinion), accuracy (whether presented information matches reality), trustworthiness (whether information appears honest and non-manipulative), credibility (perceived expertise and authority of the source), coverage legitimacy (whether coverage appears to originate from an established weather information source), and presenter legitimacy (overall perceived legitimacy of the communicator). Each item used a five-point scale with written, dimension-specific anchors.

The two legitimacy items shared identical operational definitions in the instrument; the distinction between them was carried by the question stem (coverage versus presenter). Because each dimension was measured using a single item rather than a multi-item scale, dimensional structure was assessed using exploratory factor analysis rather than internal consistency coefficients.

Participants also ranked all four sources on three practical criteria: clarity of information presentation, effectiveness of visual information, and usefulness for safety decisions. Rankings were forced-choice with no ties permitted (1 = most favorable). All clips and rating items were randomized in Qualtrics to reduce order effects.

### Stimulus Materials

Stimuli consisted of approximately three-minute clips of tornado coverage drawn from a single severe weather outbreak on April 27, 2024 affecting the Oklahoma City metropolitan region. Broadcast clips were drawn from two major-market Oklahoma City stations selected for their standardized broadcast production conventions and availability on YouTube. The event was selected because comparable coverage from both broadcast stations and YouTube weatherfluencers was available for roughly the same storm period.

YouTube clips were selected from two weatherfluencers with distinct credential profiles: Creator A, with approximately 2 million subscribers at the time of data collection; Creator B, with approximately 175,000 subscribers at the time of data collection. All four clips included comparable informational elements and were drawn from similar points in the outbreak timeline to control for event context. Native station and channel branding remained visible. Participants were not informed of presenters' professional credentials; judgments about expertise were based solely on the clips.

### Interview Protocol

Interview sessions included unobserved time for survey completion, optional think-aloud video review, and semi-structured discussion. Participants were asked to verbalize impressions while re-watching selected clips and to reflect on credibility cues, presentation features, and how they would use the information for weather-related decision-making.

### Analytical Approach

For quantitative analysis, participant-level platform means were computed by averaging ratings for the two sources within each platform, yielding paired observations for each evaluative dimension. Platform differences were tested using paired-samples t-tests with effect sizes reported as Cohen's dz. Benjamini–Hochberg false discovery rate correction ($q = .05$) was applied across platform comparisons. Wilcoxon signed-rank tests were conducted as nonparametric robustness checks, and participant fixed-effects OLS models with cluster-robust standard errors were estimated as an additional sensitivity test.

Within-platform presenter comparisons used paired t-tests with BH-FDR correction and Wilcoxon robustness checks. Ranking data were analyzed using Friedman omnibus tests followed by pairwise Wilcoxon post-hoc comparisons. To assess dimensional structure, exploratory factor analysis (KMO, Bartlett's test, varimax rotation) was conducted on participant-level mean ratings.

Qualitative data were analyzed using reflexive thematic analysis (Braun & Clarke, 2012). A single researcher conducted iterative open coding across all eight transcripts, refining the code structure through repeated engagement with the data and memoing. Qualitative accounts were systematically compared against each participant's quantitative ratings to support analytic triangulation.

## FINDINGS

Our analysis shows how viewers evaluate weather information across traditional broadcast and digital streaming platforms, addressing how trust, credibility, and legitimacy shape severe weather sensemaking and decision-making. We present findings in two parts: quantitative survey results (n=58) showing systematic evaluation patterns across six evaluative dimensions and three practical utility measures, followed by qualitative interview findings (n=8) showing the mechanisms underlying these patterns and revealing how contextual factors shape information processing during severe weather events.

### Survey Findings

Our analysis of 58 participants reveals significant and persistent advantages for traditional television meteorology over YouTube-based livestreamers across multiple evaluative dimensions.

#### *Data Quality & Measurement Validation*

After removing incomplete or invalid responses, the final quantitative sample included 58 participants. Exploratory factor analysis confirmed data suitability (KMO = .901; Bartlett's $p < .001$) and identified a dominant first factor (eigenvalue = 4.858) consistent with a strong global credibility signal, with a two-factor solution separating legitimacy items from the remaining dimensions, supporting their analytical distinctiveness.

#### *Platform-Level Differences: TV vs. YouTube*

Broadcast television meteorologists were rated higher than YouTube weatherfluencers on five of six evaluative dimensions: accuracy, trustworthiness, credibility, coverage legitimacy, and presenter legitimacy (all BH-FDR $p \leq .004$; Table 2).

**Table 2. Platform Differences in Weather Coverage Evaluation**
***p < .05, **p < .01, ***p < .001**

| Dimension | YouTube M(SD) | TV M(SD) | Δ | Cohen's *dz* | *t* | *p*FDR |
|---|---|---|---|---|---|---|
| Objectivity | 3.69 (0.83) | 3.89 (0.73) | +0.20 | 0.226 | 1.719 | .091 |
| Accuracy | 3.62 (0.69) | 4.02 (0.65) | +0.40 | 0.534 | 4.070 | <.001*** |
| Trustworthiness | 3.51 (0.79) | 3.84 (0.68) | +0.33 | 0.408 | 3.106 | .004** |
| Credibility | 3.54 (0.85) | 3.97 (0.62) | +0.43 | 0.501 | 3.814 | <.001*** |
| Coverage Legitimacy | 3.42 (0.90) | 3.97 (0.71) | +0.54 | 0.632 | 4.810 | <.001*** |
| Presenter Legitimacy | 3.49 (0.88) | 4.01 (0.69) | +0.52 | 0.596 | 4.536 | <.001*** |

BH-FDR correction applied across all six dimensions as a single family (q = .05). SDs reflect variability of participant-level platform means. Wilcoxon signed-rank robustness checks confirmed all parametric results.

The largest disparities emerged in legitimacy evaluations, where television demonstrated medium advantages for both coverage legitimacy (dz = 0.63) and presenter legitimacy (dz = 0.60). Objectivity was the sole dimension where platforms achieved statistical parity (dz = 0.23, BH-FDR p = .091), suggesting that platform-native creators can match broadcast meteorologists in perceived neutrality of presentation without formal credentialing.

#### *Within-Platform Variation: Individual Presenter Effects*

Within the YouTube platform, Creator B outperformed Creator A on five of six dimensions after BH-FDR correction, with effect sizes approaching the magnitude of platform-level TV–YouTube differences.

**Table 3. Within-Platform Presenter Comparison (YouTube)**

| Dimension | Creator A M(SD) | Creator B M(SD) | Δ | Cohen's *dz* | *p*FDR |
|---|---|---|---|---|---|
| Objectivity | 3.60 (0.97) | 3.78 (1.06) | +0.17 | 0.171 | .273 |
| Accuracy | 3.38 (0.97) | 3.86 (0.89) | +0.48 | 0.524 | .011* |
| Trustworthiness | 3.31 (1.08) | 3.71 (0.96) | +0.40 | 0.392 | .027* |
| Credibility | 3.34 (1.02) | 3.74 (1.00) | +0.40 | 0.396 | .011* |
| Coverage Legitimacy | 3.22 (1.08) | 3.62 (1.02) | +0.40 | 0.381 | .011* |
| Presenter Legitimacy | 3.22 (1.04) | 3.76 (1.05) | +0.53 | 0.515 | .004** |

BH-FDR correction applied within the YouTube family of six tests (q = .05). Wilcoxon signed-rank robustness checks confirmed all parametric results. *p < .05, **p < .01. These findings suggest that YouTube weatherfluencers may be able to partially offset platform-level credibility disadvantages through individual credential signals.

### *Coverage Legitimacy vs. Presenter Legitimacy: Within-Source Patterns*

Our analysis distinguished between two legitimacy constructs: coverage legitimacy (evaluation of the information itself) and presenter legitimacy (evaluation of the individual delivering information). Within-source correlations were strong across all four sources (r = .66–.80), indicating that content-level and person-level legitimacy judgments moved together in practice.

**Table 4. Coverage vs. Presenter Legitimacy by Source**

| **Source** | **Coverage Legitimacy M(SD)** | **Presenter Legitimacy M(SD)** | **Δ** | **pFDR** | **Correlation (r)** |
|---|---|---|---|---|---|
| Creator A | 3.22 (1.08) | 3.22 (1.04) | 0.0 | 1.000 | 0.77 |
| Creator B | 3.62 (1.02) | 3.76 (1.05) | +0.14 | .527 | 0.78 |
| CH9 | 3.78 (1.01) | 3.79 (1.12) | +0.02 | 1.000 | 0.66 |
| KFOR | 4.16 (0.87) | 4.22 (0.77) | +0.07 | .643 | 0.80 |

No source showed a statistically significant difference between the two legitimacy items after BH-FDR correction (all pFDR ≥ .527). Rather than distinguishing the information from the communicator delivering it, participants appear to have evaluated source legitimacy as a bundled judgment, consistent with the measurement constraint noted above, where both items shared identical embedded definitions.

### *Performance Rankings Reinforce Platform Hierarchies*

Participants ranked all four sources across three practical criteria: clarity of presentation, visual effectiveness, and usefulness for safety decision-making. Lower scores equal better rankings. Friedman omnibus tests confirmed significant rank differences across all four sources before platform contrasts were examined (Clarity: $\chi^2 = 12.46$, p = .006; Visual Effectiveness: $\chi^2 = 13.37$, p = .004; Safety Utility: $\chi^2 = 18.27$, p < .001). Television sources demonstrated consistent advantages across all three criteria after BH-FDR correction (q = .05; identical adjusted p = .031 for all three criteria reflects correction applied to closely ranked raw values of .016, .024, and .031).

**Table 5. Platform Ranking Performance**

| **Dimension** | **YouTube Rank (Mean)** | **TV Rank (Mean)** | **Δ** | **W** | ***p*FDR** |
|---|---|---|---|---|---|
| Clarity | 2.71 | 2.29 | -0.41 | 197.0 | .031* |
| Visual Effectiveness | 2.69 | 2.31 | -0.38 | 247.0 | .031* |
| Safety Utility | 2.69 | 2.31 | -0.38 | 241.5 | .031* |

Paired Wilcoxon signed-rank tests on participant-level platform mean ranks. BH-FDR correction applied across three ranking criteria (q = .05). *p < .05.

Pairwise posthoc comparisons showed that Creator A ranked significantly lower than all other sources across all three criteria (all BH-FDR $p \leq .021$), while Creator B was statistically indistinguishable from both television sources on all three criteria. This pattern suggests that individual presenter characteristics within YouTube can substantially offset platform-level ranking disadvantages.

### Interview Findings

Eight participants completed follow-up interviews after the survey. Four were categorized as weather-knowledgeable (P2, P3, P6, and P8), and four reflected general-public perspectives. Most participants (6/8) were unfamiliar with YouTube weather livestreamers prior to the study. The interviews help explain the survey patterns by showing how participants used platform, visual, and communicative cues to judge trust, credibility, legitimacy, and practical usefulness across formats.

#### *Official Authority and Professional Gatekeeping*

Participants consistently treated broadcast affiliation as a credibility signal because it implied vetting, formal credentials, and accountability. Television meteorologists were assumed to have passed through rigorous selection processes, whereas YouTube was treated as an open-entry environment in which anyone can start a stream. As P6 explained, *"Being associated with a news station to me instantly made me feel they were more credible, because I know kind of what goes into selecting people...whereas anyone can kind of start a stream."* P3 articulated a similar assumption about formal qualification requirements: *"In order to get a position as a TV meteorologist, you have to have a degree in meteorology or atmospheric science...versus on YouTube, anybody can start a YouTube channel."*

Participants also associated broadcast credibility with broader accountability structures. Traditional television was understood as more regulated and therefore less open to low-quality or irresponsible communication. By contrast, YouTube's lack of equivalent gatekeeping was seen as a source of uncertainty about quality control. At the same time, participants suggested that legitimacy was not fixed: some noted that endorsement from professional meteorologists or official forecasters could increase confidence in an independent creator. This suggests that legitimacy was tied not only to platform or format, but to whether a communicator could be linked to recognized professional standards.

#### *Platform Familiarity and Expectation Effects*

Participants often brought platform-based expectations to their evaluations before closely assessing the content itself. YouTube creators were frequently associated with monetization, attention-seeking, or entertainment incentives, while television meteorologists were more often assumed to be oriented toward public safety and local service.

P5 described one creator as potentially *"just trying to like, induce fear, and be like 'this is the end of the world' and news-y. So it's good for his channel. I mean, he's probably getting views and [is getting] money."* P6 made the contrast even more explicit: *"on the broadcast side, a lot of times, from what I've seen, the motivation is very much safety. Protect people. Alert the local communities, whereas on the YouTuber side, that makes me think that they're more there for providing entertainment, which turns into monetization."*

These assumptions mattered even when participants tried to resist them. P5 explicitly stated, *"I wouldn't by default, choose like TV media just over the other ones. Again…I would just watch a few to see videos to see how legitimate and how accurate they are."* Yet even this participant said they would be more likely to seek out a news outlet on YouTube than an independent creator. In other words, platform expectations operated both as explicit evaluative criteria and as background assumptions shaping information-seeking behavior.

#### *Professional Presentation and Surface Credibility Signals*

Production quality and physical setting also shaped credibility judgments. Participants quickly used visual details as proxies for professionalism. Informal or domestic-looking environments reduced confidence for several participants, especially when contrasted with conventional broadcast settings. P4 stated, *"The second guy had his background on his bedroom or something, so definitely not credible."*

Participants also compared production professionalism within YouTube itself. Creators who used more structured setups or higher-quality equipment were sometimes seen as more serious communicators. P7 contrasted *"the YouTuber who seems to like have his bedroom in the background, and not the YouTuber, who has, like the* *professional mic, set up,"* suggesting that production choices influenced initial judgments even within the same platform.

However, participants emphasized that visual presentation alone did not determine credibility. Professional attire or polished appearance could not compensate for weak explanations or unconvincing delivery. P4 noted that one presenter wore a suit but still seemed untrustworthy because "*it was more of like his body language and what he was saying."* While surface cues shaped first impressions, participants generally treated them as secondary to communication quality and meteorological reasoning.

#### *Technical Competency and Communication Effectiveness*

Technical accuracy served as a fundamental credibility requirement across participants. Errors involving warning interpretation quickly undermined trust. One creator incorrectly stated that a tornado was in progress when only a severe thunderstorm warning was active; P6 noted that *"this mistake kinda instantly put me off a little bit."*

Participants also evaluated how presenters communicated technical information. P2 appreciated when presenters could seamlessly transition between jargon and plain language. Communication clarity emerged as a key difference between formats. P6 observed that television meteorologists *"explained things a little bit more …accessible to audiences without meteorological backgrounds. Whereas the streamers very much seem to be focusing on a lot of that technical jargon."*

Participants also valued explanations that connected current observations to likely outcomes—for example, clarifying whether rotation indicated an imminent tornado or explaining how radar patterns related to storm behavior.

#### *Entertainment vs. Information Tension*

Participants repeatedly distinguished between communication styles that felt safety-oriented and those that felt attention-oriented. P7 reacted strongly to one creator's delivery style, saying, *"his voice was trying to get my attention…you use that kind of voice to get into people's dopamine systems and get them to really engage with your content."* Their reaction was visceral: *"I resent the fact that he's using this voice. I resent his voice right now. It feels like very manipulative."*

Other participants also linked spectacle to misplaced priorities. P2 suggested that dramatic tornado footage was likely *"what's drawing [a lot of] clicks,"* and P8 captured the tension from the perspective of a person facing actual risk: *"And it's like they're having fun talking about this massive tornado outbreak. And like using all these words that they learn. And it's like, okay. Well, I'm trying to live."* Visible monetization cues further intensified this concern. P6 noted that when they saw sponsorship signals, it brought the presentation down *"a little bit…for entertainment."*

These concerns were not limited entirely to YouTube. Participants also criticized excessive energy or anxiety-inducing delivery in television coverage. P7 described one television meteorologist as making them feel *"anxious and uncomfortable."* Still, participants more often interpreted television as operating within a public-service framework, even as they acknowledged that ratings pressures could exist there. The more salient distinction was not platform but whether a presenter's style seemed calibrated toward helping viewers act under pressure or toward sustaining their engagement.

#### *Temporal Reasoning and Prediction Needs*

Participants valued coverage that situated current conditions within broader temporal narratives of storm evolution and trajectory. P1 noted the usefulness of coverage showing *"figures in different [places] separated in different times… they [presented] figures at 2:40pm. And then we talk about the 3:40pm. Maybe 1h later…and then [they] provided some suggestions [like] what's the weather going on in next few hours?"*

Weather-knowledgeable participants also valued explanations linking current observations to likely outcomes. P8 appreciated presentations that explained *"this is a confirmed tornado that we saw on the ground, or this is rotating, and this kind of pattern often produces tornadoes."*

Participants also perceived differences between formats. P4 observed that television broadcasters *"were more focused on the broader picture. What's coming ahead,"* while streamers were *"more focused on what's going on right now."*

#### *Place-Based Expertise and Geographic Context*

Participants also discussed the importance of geographic familiarity when interpreting storms. P8 emphasized wanting communicators who understood local terrain and locations: *"I want somebody to actually know where I am to know these areas, know which cities, know how to explain to me the path of a storm."*

Weather-knowledgeable participants similarly described local expertise as an advantage for broadcast meteorologists. P6 noted that local meteorologists *"know their local areas…they'll know different social vulnerabilities of different areas"* whereas streamers might not be located where storms occur.

However, this expectation was not universal. P5 argued that geographic familiarity mattered less than interpretive skill, stating that credibility depended on *"how they process the data that they get and how [they] process that…it's just a matter of how they're reading the data."*

## DISCUSSION

### Popularity without Legitimacy?

The trajectory of the two YouTube creators in this study is illustrative. In April 2024, Creator A had approximately 2 million subscribers and Creator B approximately 175,000. By early 2026, those figures had grown to approximately 3.1 million and 1.7 million, respectively. This growth, alongside persistent legitimacy concerns, is difficult to reconcile with existing crisis communication models.

Prior crisis informatics research offers a useful starting point. Chauhan and Hughes' work on Crisis Named Resources (CNRs) documents how such resources can accumulate visibility and audience engagement during crisis events despite unknown operators and uncertain credibility (Chauhan & Hughes, 2018, 2020, 2021). Related work on digital volunteers likewise shows how non-official actors can become consequential in the circulation and interpretation of crisis information (Palen et al., 2009). In both cases, visibility emerges through structural features of the information environment rather than professional credentialing.

The weatherfluencer case shares this essential feature but departs from the CNR model. CNRs are asynchronous aggregators: Twitter accounts, Facebook pages, or Reddit threads named after a crisis event that collect and circulate information episodically. They derive their visibility from event-specific naming and crisis-time discoverability, and they typically disappear when the event does. Weatherfluencers are fundamentally different in both modality and temporality. They are synchronous performers engaged in continuous, real-time narration of unfolding conditions to a live audience cultivated across many events over time. Where CNRs are relatively transient, weatherfluencers persist across events, with their subscriber networks and platform visibility outlasting any individual event. Their audience growth reflects a durable structural feature of the contemporary severe weather information environment rather than a temporary anomaly.

This persistence also invites reconsideration of how trust is produced in platform environments. Wachinger et al. (2013) demonstrate that trust in risk-related sources is socially produced and historically grounded. Platform environments generate their own forms of social validation, including metrics such as subscriber counts and view visibility that can function as trust proxies without carrying the institutional accountability those signals were historically bundled with. Our participants demonstrated an informal awareness of this distinction: they recognized weatherfluencers as part of their information environment while simultaneously discounting them for safety-critical decisions, suggesting a separation between platform visibility and professional authorization.

### Legitimacy and Attention Dynamics

The broadcast advantage on legitimacy-related dimensions is consistent with classic warning response models, which conceptualize trust and source credibility as relationships between publics and official warning systems that support attention, comprehension, and protective action (Lindell & Perry, 2012; Mileti & Sorensen, 1990). These signals functioned as durable heuristics for appropriate authority, consistent with weather-specific research showing that affiliation shapes how people interpret warnings, assess risk, and decide whether to take protective action (Ripberger et al., 2015; Sherman-Morris, 2005). That the largest empirical advantages for broadcast sources emerged specifically on the two legitimacy dimensions, rather than on objectivity or accuracy, further underscores that what participants were responding to was not simply perceived informational quality but the information scaffolding within which that information was delivered.

What the existing literature does not fully anticipate, however, is the pattern that emerges on the attention side of the equation. Crisis informatics research has shown that credibility judgments during crisis events emerge through processes of verification, correction, and collective sensemaking rather than through stable deference to authorities (Mendoza et al., 2010; Starbird et al., 2014). This study extends that insight in a specific direction: even when audiences produce clear and consistent legitimacy judgments that favor official sources, those judgments do not straightforwardly determine which communicators attract and retain attention. Weatherfluencers remained salient to participants and were recognized as meaningful parts of the severe weather information landscape despite being evaluated as less legitimate and less appropriate for safety-critical guidance. Related work on data-driven hazard communication highlights real-time warning improvement as a priority for public risk understanding, with implications for livestream-based weather communication (Saunders et al., 2025).

### Weatherfluencers in Platformized Environments

Understanding how weatherfluencers sustain attention despite legitimacy deficits requires examining the communicative affordances of livestreaming. HCI and communication research shows that audiences rely on visible surface cues, such as presentation style, perceived expertise, and interactivity, when evaluating online information and its source (Fogg, 2003; Metzger, 2007). In broadcast contexts, these surface cues are bundled with credibility signals: studio production, graphics, and station branding that visibly index organizational affiliation. Platform environments disaggregate this bundle. A weatherfluencer may display production quality and technical competence that would signal professional standing in broadcast contexts, but without the organizational affiliation participants treated as the basis for legitimacy. Our qualitative findings show participants working through this disaggregation, noting production quality while questioning motivation and accountability. This need to separate surface credibility from official authorization is not well captured by credibility frameworks developed for environments where those cues reliably co-occurred.

The livestream modality intensifies this dynamic through performative presence. Unlike the asynchronous posts characteristic of CNRs, livestreaming places the communicator in continuous co-presence with their audience during an unfolding event. The weatherfluencer functions less like a crisis-named social media account and more like a sports commentator, whose interpretive authority is performed through demonstrated competence, sustained engagement, and real-time responsiveness rather than through professional credential or organizational mandate. Our participants described something analogous: weatherfluencers were evaluated as experientially immediate and continuously present even when explicitly discounted as authoritative sources for protective action. This dynamic highlights relational and experiential dimensions of crisis interpretation not captured by traditional credibility frameworks.

This relational dynamic is further shaped by the monetization structures that platform livestreaming makes visible. Hurcombe (2025) identifies platformatization as a defining dimension of newsfluencers, noting that platform revenue streams are not incidental to content production but structurally embedded within it, with recommendation algorithms creating pressure toward attention-maximizing presentation that can tension with professional judgment. In the livestream context, this monetization is not separated from the broadcast as it is in traditional television; it is visible within the stream itself. YouTube's Super Chat feature allows viewers to pay to have their messages highlighted during a live broadcast (YouTube, 2026). Several participants in this study noted visible commercialization signals as credibility-reducing cues, consistent with Hurcombe's observation that the economic logics of platform content creation can conflict with norms of accuracy and impartiality. This monetization visibility contrasts with broadcast television, where advertising is separated from editorial content. The result is an information environment in which the financial incentives shaping a communicator's presentation choices are simultaneously more visible to audiences and more difficult to evaluate than those operating within broadcast contexts.

### Format Affordances and Sensemaking

The legitimacy-attention gap documented in this study is not only a product of platform structures and monetization logics; it also reflects a functional differentiation between broadcast and livestreaming formats in how they support audience sensemaking during unfolding severe weather events. Broadcast meteorology has historically organized severe weather communication around episodic, trajectory-oriented coverage that translates technical meteorological products into locally meaningful, place-embedded guidance through professional conventions and sustained community presence (Henson, 2013). Livestreaming, as a format, is structurally oriented toward different sensemaking needs: continuous real-time narration, sustained co-presence during uncertainty, and immediate interpretive responsiveness to rapidly changing conditions.

This functional differentiation has implications for how authority and format affordances interact. The place-based interpretive authority that Henson (2013) identifies as central to broadcast meteorology's social function is structurally tied to the historical context of local broadcasting, through community embeddedness and geographic accountability. Platform-native livestreaming assembles audiences through algorithmic recommendation and subscriber networks rather than geographic proximity, producing a different kind of interpretive relationship between communicator and audience. That relationship, as Hurcombe (2025) observes of newsfluencers more broadly, is characterized by parasocial intimacy and relational labor rather than by the community accountability that grounds local broadcast authority. These are structurally different bases for interpretive trust, even when the meteorological content being delivered is comparable.

Broadcast meteorologists increasingly maintain online presences through station websites, mobile applications, and official social media accounts. However, station-based digital platforms prioritize curated updates and episodic access consistent with established broadcast communication routines, while platform-native livestreaming foregrounds continuity and real-time co-presence in ways that legacy digital channels do not structurally replicate. Whether broadcast meteorologists adopting livestreaming formats would close the sensemaking gap suggested by this study, or whether the format advantages associated with weatherfluencers are inseparable from the platform contexts and relational dynamics in which they operate, remains an open empirical question and a productive direction for future research.

## IMPLICATIONS

### Complementary Formats Within Distinct Infrastructures

These findings do not suggest that broadcast meteorologists abandon formal professional norms. Rather, they invite reflection on how distinct communication infrastructures can be understood as complementary rather than competing resources for public sensemaking during severe weather events.

For broadcast meteorologists and the news organizations that employ them, professional legitimacy remains legible and consequential to audiences even in platform-saturated information environments. Credentialing, organizational accountability, and local community embeddedness continue to function as meaningful signals of appropriate authority for safety-critical guidance. Broadcast meteorologists operating on platform-native channels may need to make their professional affiliations and credentials explicitly visible in ways that broadcast contexts make automatic, given that platform environments structurally resemble open-entry creator ecosystems where such signals are not assumed. The format-level sensemaking gap identified by participants also suggests a potential avenue for station-based broadcasters to consider how livestreaming might extend their public service function during active severe weather events, particularly during rapidly evolving conditions between scheduled broadcasts.

For emergency management agencies and public information officers, the legitimacy-attention relationship documented here cannot be assumed to hold in platformized information environments. Audiences may simultaneously recognize weatherfluencers as less legitimate and less appropriate for safety-critical guidance while continuing to encounter and engage with their content. Emergency communication strategies built on the assumption that trusted sources attract the most attention may need revision to account for platform environments where visibility emerges through algorithmic dynamics rather than credentialing. Agencies might productively consider how formal endorsement or collaborative relationships with credentialed platform-native communicators could extend safety guidance into communities already assembled around weatherfluencer content.

For platform-native weather communicators the findings surface a specific challenge around credential visibility and monetization transparency. Participants recognized visible commercialization signals as confidence reducing even when technical competence was apparent. Communicators who hold formal meteorological credentials may benefit from making credentials explicitly visible to audiences encountering them outside broadcast frameworks, and those who address the monetization transparency challenge may be better positioned to close the legitimacy gap that platform context otherwise produces.

These findings suggest the need for crisis communication research to more explicitly account for attention as a distinct analytic dimension alongside trust, credibility, and legitimacy. Existing crisis informatics models provide strong explanations of how people evaluate information and sources but offer more limited insight into why certain interpreters attract attention even when evaluated as less legitimate or less appropriate for safety-critical guidance. Weatherfluencers make this tension analytically visible for crisis communication research.

## LIMITATIONS

This study has several limitations that should be considered when interpreting the findings. First, while the quantitative sample size was sufficient to detect the observed effects, the interview component was intentionally small and purposively sampled. As a result, the qualitative findings illustrate evaluative mechanisms rather than exhaustively representing the full diversity of public sensemaking practices. Demographic data beyond age, region, and weather experience were not collected for the MTurk sample.

Second, the study's geographic and hazard focus constrain transferability. The controlled viewing context cannot fully capture the heightened emotional stakes, time pressure, and personal risk present during real severe weather events.

Interviews followed survey completion, which may have influenced participants' reflections and sensemaking strategies. Although this sequencing enabled direct comparison between ratings and explanations, it may differ from naturalistic media consumption during unfolding events. Additionally, this study did not examine whether official weather services such as the National Weather Service that broadcast via YouTube function equivalently to either broadcast television or independent weatherfluencers from an audience evaluation perspective.

Finally, this study focused on comparisons between television meteorologists and YouTube weather livestreamers and did not examine other influential intermediaries such as weather apps or automated alerting systems. The interview sample also did not include active followers of the YouTube creators evaluated in the study; future research that specifically recruits weatherfluencer audiences would provide important insight into why viewers choose these sources and what evaluative criteria they apply, complementing the first-impression evaluations documented here. Analysis of live chat data represents a productive direction, offering a window into how audiences collectively negotiate credibility and interpretive authority.

## CONCLUSION

Broadcast meteorologists retained consistent advantages over YouTube weatherfluencers on credibility, legitimacy, and safety utility, with the largest gaps on legitimacy. Yet, weatherfluencers remained salient parts of participants' information environments despite being evaluated as less legitimate and less appropriate for safety-critical guidance. This decoupling of audience attention from official authorization highlights a gap in crisis communication models and underscores the need to better account for how visibility, legitimacy, and interpretive authority interact in platform-based severe weather communication.

## ACKNOWLEDGMENTS

We thank our participants and the CommPrac Lab. This research was supported by a doctoral research grant from the University of Washington Department of Human Centered Design & Engineering.